\documentclass{elsart}
\usepackage{amsfonts}
\usepackage{amssymb}
\usepackage{amsmath}
\usepackage{dcolumn}
\usepackage[all]{xy}
\usepackage[]{epsfig}
\begin{document}

\begin{frontmatter}

\title{Consequences of vacuum polarization on electromagnetic waves in
a Lorentz-symmetry breaking scenario}

\author{B. Agostini\thanksref{Braulio}}
\thanks[Braulio]{e-mail address: agostini@cbpf.br}
\address{Centro Brasileiro de Pesquisas F\'\i sicas, Rio de Janeiro, RJ, Brasil}
\author{F. A. Barone\thanksref{Fabricio}}
\thanks[Fabricio]{e-mail address: fbarone@unifei.edu.br}
\address{ICE - Universidade Federal de Itajub\'a, Itajub\'a, MG, Brasil}
\author{F. E. Barone\thanksref{Barone}}
\thanks[Barone]{e-mail address: febarone@cbpf.br}
\address{Centro Brasileiro de Pesquisas F\'\i sicas, Rio de Janeiro, RJ, Brasil}
\author{Patricio Gaete\thanksref{cile}}
\thanks[cile]{e-mail address: patricio.gaete@usm.cl}
\address{Departmento de F\'{\i}sica and Centro
Cient\'{\i}fico-Tecnol\'{o}gico de Valpara\'{\i}so, Universidad
T\'ecnica Federico Santa Mar\'{\i}a, Valpara\'{\i}so, Chile}
\author{J. A. Helay\"el-Neto\thanksref{Helayel}}
\thanks[Helayel]{e-mail address: helayel@cbpf.br}
\address{Centro Brasileiro de Pesquisas F\'\i sicas, Rio de Janeiro, RJ, Brasil}

\begin{abstract}
The propagation of electromagnetic waves in a Lorentz-symmetry
violating scenario is investigated in connection with non-linear
(photon self-interacting) terms induced by quantum effects. It turns
out that the photon field acquires an interesting polarization state
and, from our calculations of phase and group velocities, we contemplate different scenarios
with physically realizable magnetic fields and identify situations where non-linearity effects dominate
over Lorentz-symmetry breaking ones and vice-versa.
\end{abstract}
\end{frontmatter}

\section{Introduction}

In the endeavor of unveiling  the physics that unfold beyond the
Standard Model, our best efforts have led us to a whole class of
CPT- and Lorentz-symmetry breaking models. The most suitable
framework to deal with questions within these scenarios is the
effective theory approach referred to as the Standard Model
Extension (SME) \cite{Kostelecky}, where it is possible to embody
spontaneous Lorentz-symmetry breaking and still keep desired
properties of standard Quantum Field Theory. Recently, a great deal
of efforts have been devoted to the search of measurable
consequences of this kind of breaking \cite{Jackiw,dados}. On the
other hand, although direct measurements of nonlinear effects of
electromagnetic processes in vacuum remain elusive, intensive
experimental research is still in course; for example, in the PVLAS
apparatus \cite{PVLAS}, now operating in its Phase II
\cite{Cantatore, Valle}. Also, we point out another interesting
class of experiments (light-shining-through-a-wall, LSW) where
non-linear effects (photon-photon-ALP coupling) are to be detected
\cite{Ringwald1, Ringwald2}.

Here, we address the issue of light propagation in a non-trivial
background. We consider the vacuum in the presence of a strong
magnetic field in a scenario with Lorentz-symmetry breaking.
Non-linearity due to quantum field-theoretic effects, on the one
hand, and Lorentz-symmetry violation induced phenomena, on the other
hand, are all very tiny at our accessible energy scales. But,
precision measurements in electromagnetic processes can be carried
out to set up constraints and bounds on physical parameters related
to these non-linear and Lorentz-symmetry violating effects, shedding
some light on a physics that becomes very significant at high
energies, at scales close to the threshold for more fundamental
theories.

Our work sets out to propose a discussion on the interplay between
non-linear and Lorentz-symmetry violation considered contemporarily.
We have clear that both effects are per s\'{e} very small and we
actually understand that their interference is much far from our
precision measurements. However, we are able to show how these
phenomena add to each other and we identify physical situations
where they can be of the same order. Also, our results are shown to
be compatible with current bounds carefully analyzed in the work of
Refs. \cite{dados,Astro} and papers quoted therein. In the present
work, by virtue of the Lorentz-symmetry violation in the presence of
a strong magnetic field in vacuum, we find out that the photon
acquires a peculiar longitudinal polarization that can be tested at
laboratory scales, possibly in a PVLAS-like configuration, for
instance. Also, the characteristic phase and group velocities, whose
difference is useful to characterize dispersion, are worked out.
Along the paper, we shall adopt the metric $\eta^{\mu\nu}=(+,-,-,-)$
and the Levi-Civita tensor defined such that $\epsilon^{0123}=1$.

\section{Model under consideration}

The model under consideration is described by the Lagrangian
density:
\begin{eqnarray}
\label{modelo}
{\cal L}=\frac{1}{8\pi}({\bf E}^{2}-{\bf
B}^{2})+\frac{R}{8\pi}({\bf E}^{2}- {\bf
B}^{2})^{2}+\frac{S}{8\pi}({\bf E}\cdot{\bf B})^{2}
+\frac{1}{16\pi}\epsilon^{\mu\nu\lambda\rho}A_{\nu}(\hat{k}_{AF})_{\mu}F_{\lambda\rho},
\end{eqnarray}
where ${\bf E}$ and ${\bf B}$ are the electric and magnetic fields,
$A^{\mu}$ is the vector potential,
$F^{\mu\nu}=\partial^{\mu}A^{\nu}-\partial^{\nu}A^{\mu}$ is the
field strength and $(\hat{k}_{AF})_{\mu}$ is the vector operator that
characterizes the Lorentz-violating Chern-Simons term proposed in the work of Ref. \cite{AM10},
\begin{eqnarray}
\label{defkaf}
(\hat{k}_{AF})_{\mu}&=&\!\!\!\sum_{d=3,5,...}(k_{AF}^{(d)})_{\mu}^{\alpha_{1},\alpha_{2},...,\alpha_{(d-3)}}\partial_{\alpha_{1}}...\partial_{\alpha_{(d-3)}}\cr\cr
&=&\!\!\!(k_{AF}^{(3)})_{\mu}+(k_{AF}^{(5)})_{\mu}^{\alpha_{1},\alpha_{2}}\partial_{\alpha_{1}}\partial_{\alpha_{2}}+
(k_{AF}^{(7)})_{\mu}^{\alpha_{1},\alpha_{2},\alpha_{3},\alpha_{4}}\partial_{\alpha_{1}}\partial_{\alpha_{2}}\partial_{\alpha_{3}}\partial_{\alpha_{4}}+... \ .
\end{eqnarray}

It is important to notice that the last term in (\ref{modelo}) along with the expression of Eq. (\ref{defkaf}) generalizes the so-called
Carroll-Field-Jackiw term \cite{Jackiw}, for which $(k_{AF}^{(3)})^{\mu}= v^{\mu}$ and $(k_{AF}^{(d)})^{\mu}=0$ for $d\geq5$ .

Taking
\begin{eqnarray}
\label{Parametros} R=\frac{e^{4}}{45\pi m_{e}^{4}}\ ,\ S=7R\ ,
\end{eqnarray}
where $m_{e}$ is the electron mass, and, for $(\hat{k}_{AF})_{\mu}=0$, we are led
to the well-known Euler-Heisenberg Lagrangian \cite{Schwinger51}.
This describes an effective model up to first order in the
parameters $R$ and $S$; so, our calculations shall be carried out
only up to first order in $R$ and $S$. We stress that we discard
terms of order $Rv^{\mu}$ and $Sv^{\mu}$, since $v^{\mu}$ is very
tiny according to the available experimental data \cite{dados}.We
also point out that our (Lorentz- symmetry violating) Chern-Simons
term in the Lagrangian (\ref{modelo}) replaces the P- and
CP-violating term, $\left( {{\bf B}^2  - {\bf E}^2 } \right){\bf E}
\cdot {\bf B}$, considered by Hu and Liao in the paper of Ref.
\cite{Hu}. We adopt the choice of the Chern-Simons term to describe
CPT violation and, then, to study its consequence in presence of the
non-linear $ \left( {{\bf B}^2  - {\bf E}^2 } \right)$ and $\left(
{{\bf E} \cdot {\bf B}} \right)^2$ terms.

From now on, we shall always restric to field configurations where the effects imposed by the non-linear terms as well as the ones imposed the Chern-Simons-like term in (\ref{modelo}) are small perturbations.

In order to bring the dynamical equations into a compact and
convenient form, we shall define the vectors ${\bf D}$ and ${\bf
H}$, in analogy to the electric displacement and magnetic field
strength, as follows:
\begin{eqnarray}
{\bf D}=4\pi\frac{\partial}{\partial{\bf E}}\Biggl[
\frac{1}{8\pi}({\bf E}^{2}-{\bf B}^{2})+\frac{R}{8\pi}({\bf E}^{2}-
{\bf B}^{2})^{2}+\frac{S}{8\pi}({\bf E}\cdot{\bf B})^{2}\Biggr]\
,\cr\cr {\bf H}=-4\pi\frac{\partial}{\partial{\bf
B}}\Biggl[\frac{1}{8\pi}({\bf E}^{2}- {\bf
B}^{2})+\frac{R}{8\pi}({\bf E}^{2}-{\bf B}^{2})^{2}
+\frac{S}{8\pi}({\bf E}\cdot{\bf B})^{2}\Biggr]\ ,
\end{eqnarray}
so that,
\begin{eqnarray}
\label{DH} {\bf D}&=&{\bf E}+2R({\bf E}^{2}-{\bf B}^{2}){\bf
E}+S({\bf E}\cdot{\bf B}){\bf B}\ ,\cr\cr {\bf H}&=&{\bf B}+2R({\bf
E}^{2}-{\bf B}^{2}){\bf B}-S({\bf E}\cdot{\bf B}){\bf E}\ .
\end{eqnarray}
The usual methods used to study light propagation in a non-trivial
vacuum, due to the presence of external fields or boundary
conditions, can be applied here
\cite{Adler,Birula,BrezinItzykson,DittrichGiess,Kruglov,Scharnhorst,Barton,BaroneFarina}.

From now on, we shall be considering the propagation of
electromagnetic waves in the presence of an external constant and
uniform magnetic field. The goal is to identify the polarization
states of the photon field in this situation. To this end, let us
denote the electric and magnetic fields of the propagating wave by
${\bf E}_{P}$ and ${\bf B}_{P}$ respectively, and the external
magnetic field by ${\bf B}_{0}$.

The dynamical equations that stem from the Lagrangian density
(\ref{modelo}) read as below:
\begin{eqnarray}
\label{eqs} {\bf\nabla}\cdot{\bf D}={\hat{\bf k}}_{AF}\cdot{\bf B}\ \ &,&\ \
{\bf\nabla}\cdot{\bf B}=0\ ,\cr\cr {\bf\nabla}\times{\bf
E}&=&-\frac{\partial{\bf B}}{\partial t}\ ,\cr\cr
{\bf\nabla}\times{\bf H}=\frac{\partial{\bf D}}{\partial t}\!\!\! &+&\!\!\!
(\hat{k}_{AF})^{0} {\bf B}-(\hat{\bf k}_{AF})\times{\bf E}\ .
\end{eqnarray}
The simplest solutions, written in terms of the propagating and
external fields, are of the form:
\begin{equation}
\label{asd1} {\bf E}={\bf E}_{P}+\delta{\bf E}\ \ \ ,\ \ \ {\bf
B}={\bf B}_{0}+{\bf B}_{P}+ \delta{\bf B}\ ,
\end{equation}
where $\delta{\bf E}$ and $\delta{\bf B}$ stand for the
contributions that do not propagate, as dictated by the set of
equations (\ref{eqs}).

In view of what has been exposed above, the following conditions may
be adopted:
\begin{eqnarray}
\label{rel} \left|{\delta{\bf E}}\right| << \left| {{\bf E}_{P}}
\right|<<(R)^{-1/2},(S)^{-1/2} ,\nonumber\\
\left| {\delta{\bf B}} \right|<<\left| {{\bf B}_{P}} \right|<<
\left| { {\bf B}_{0}} \right|<<(R)^{-1/2},(S)^{-1/2}\ .
\end{eqnarray}
Substituting (\ref{asd1}) into (\ref{DH}), taking into account the
relations (\ref{rel}) and discarding terms of order $R\delta{\bf
E}$, $S\delta{\bf E}$, $R\delta{\bf B}$, $S\delta{\bf B}$, ${\bf
E}_{P}^{2}$, ${\bf B}_{P}^{2}$ and ${\bf E}_{P} \cdot{\bf B}_{P}$,
we can write that
\begin{eqnarray}
\label{asd2} {\bf D}={\bf E}_{P}+\delta{\bf E}-2RB_{0}^{2}{\bf
E}_{P}+S({\bf E}_{p} \cdot{\bf B}_{0}){\bf B}_{0}\ ,\cr\cr {\bf
H}={\bf B}_{0}+{\bf B}_{P}+\delta{\bf B}-2RB_{0}^{2}{\bf B}_{P}-
4R({\bf B}_{0}\cdot{\bf B}_{P}){\bf B}_{0}-2RB_{0}^{2}{\bf B}_{0}\ .
\end{eqnarray}
From this, it can be readily seen that the displacement field and
the magnetic field strength for the propagating fields can be
expressed by
\begin{eqnarray}
\label{defDHP} {\bf D}_{P}&=&{\bf E}_{P}-2RB_{0}^{2}{\bf
E}_{P}+S({\bf B}_{0}\cdot{\bf E}_{P}) {\bf B}_{0}\ ,\cr\cr {\bf
H}_{P}&=&{\bf B}_{P}-2RB_{0}^{2}{\bf B}_{P}-4R({\bf B}_{0}\cdot{\bf
B}_{P}) {\bf B}_{0}\ .
\end{eqnarray}

Replacing (\ref{asd1}) and (\ref{asd2}) into (\ref{eqs}) yields two
sets of equations, one for the non-propagating fields and the
another for the propagating ones, whose dynamics we are interested
in. We then have:
\begin{eqnarray}
\label{eqprop} {\bf\nabla}\cdot{\bf D}_{P}={\hat{\bf k}}_{AF}\cdot{\bf B}_{P}\
\ &,&\ \ {\bf\nabla}\cdot{\bf B}_{P}=0\ ,\cr\cr
{\bf\nabla}\times{\bf E}_{P}&=&-\frac{\partial{\bf B}_{P}}{\partial
t}\ ,\cr\cr {\bf\nabla}\times{\bf H}_{P}=\frac{\partial{\bf
D}_{P}}{\partial t}\!\!\!  &+&\!\!\!   (\hat{k}_{AF})^{0}\ {\bf B}_{P}-(\hat{\bf k}_{AF})\times{\bf
E}_{P}\ .
\end{eqnarray}
The explicit form of the Euclidean tensors for electric permittivity
and inverse magnetic permeability in these equations can be obtained
from (\ref{defDHP}),
\begin{eqnarray}
\label{defespilonmu}
\varepsilon_{P}^{ij}&=&(1-2RB_{0}^{2})\delta^{ij}+SB_{0}^{i}B_{0}^{j}\
,\cr\cr
(\mu_{P}^{-1})^{ij}&=&(1-2RB_{0}^{2})\delta^{ij}-4RB_{0}^{i}B_{0}^{j}\
,
\end{eqnarray}
where $\delta^{ij}$ stands for the Kronecker delta. Therefore, in
components, Eqs (\ref{defDHP}) reads as below:
\begin{eqnarray}
\label{DHP} D_{P}^{i}=\sum_{j=1}^{3}\varepsilon_{P}^{ij}E_{P}^{j}\ \
\ ,\ \ \ H_{P}^{i}=\sum_{j=1}^{3}(\mu_{P}^{-1})^{ij}B_{P}^{j}\ .
\end{eqnarray}

As usual, let us search for plane waves configurations for the
propagating fields, ${\bf E}_{P}$ and ${\bf B}_{P}$, in the form
\begin{eqnarray}
\label{propsol}
{\bf E}_{P}&=&{\bf e}\ \exp[i({\bf k}\cdot{\bf
r}-\omega t)]\ ,\cr\cr {\bf B}_{P}&=&{\bf b}\ \exp[i({\bf
k}\cdot{\bf r}-\omega t)]\ ,
\end{eqnarray}
where ${\bf e}$ and ${\bf b}$ are constant and uniform vectors. For
simplicity, we restrict our attention to the case where the wave
vector $\bf k$ is perpendicular to the external magnetic field,
${\bf B}_{0}$, i.e. ${\bf k}\cdot{\bf B}_{0}=0$, and the coordinate
system will be taken in such a way that
\begin{eqnarray}
{\bf B}_{0}=B_{0}{\hat x}\ \ \ ,\ \ \ {\bf k}=k{\hat z}\ .
\end{eqnarray}
In such a system, the tensors in (\ref{defespilonmu}) can be brought
into the form
\begin{eqnarray}
\label{zxc1}
\varepsilon_{P}^{ij}&=&(1-2RB_{0}^{2})\delta^{ij}+SB_{0}^{2}\delta^{i1}\delta^{j1}\
,\cr\cr
(\mu_{P}^{-1})^{ij}&=&(1-2RB_{0}^{2})\delta^{ij}-4RB_{0}^{2}\delta^{i1}\delta^{j1}\
,
\end{eqnarray}
and the differential operators $(\hat{k}_{AF})^{0}$ and $(\hat{\bf k}_{AF})$ operating on the fields (\ref{propsol}) can be substituted by the vectors
\begin{eqnarray}
\label{xcv1}
(\hat{k}_{AF})^{0}\rightarrow(k_{AF})^{0}&=& \sum_{d=3,5,...}(k_{AF}^{(d)})^{0}_{\ \ \alpha_{1},\alpha_{2},...,\alpha_{(d-3)}}k^{\alpha_{1}}...k^{\alpha_{(d-3)}}\cr\cr
(\hat{\bf k}_{AF}^{(d)})\rightarrow({\bf k}_{AF})&=&
\sum_{d=3,5,...}({\bf k}_{AF}^{(d)})_{\alpha_{1},\alpha_{2},...,\alpha_{(d-3)}}k^{\alpha_{1}}...k^{\alpha_{(d-3)}}\ ,
\end{eqnarray}
where the indexes $\alpha_{i}$ can assume the values $0$ or $3$, once our wave vector is taken to be given by $k^{\mu}=(\omega,0,0,k)$.

The Lorentz-breaking term as well as the non-linear terms modify the dispersion relation. In spite of this fact, we can use the Maxwell dispersion relation, where $\omega=k$, and set $k^{\alpha_{i}}=k^{0}=k^{3}=\omega=k$ in (\ref{xcv1}) because the coefficients $(k_{AF}^{(d)})^{0}_{\ \alpha_{1},\alpha_{2},...,\alpha_{(d-3)}}$ are small quantities and the wave vector $k^{\mu}$ is not large \footnote{If the wave vector becomes large, the Lagrangian (\ref{modelo}) must include terms in higher orders in the electromagnetic fields}. So, we can write expression (\ref{xcv1}) in the form
\begin{eqnarray}
\label{xcv2}
(\hat{k}_{AF})^{0}\rightarrow(k_{AF})^{0}&=& \sum_{d=3,5,...}(k_{AF}^{(d)})^{0}_{\ \alpha_{1},\alpha_{2},...,\alpha_{(d-3)}}k^{\alpha_{1}}...k^{\alpha_{(d-3)}}\cr\cr
&=&\sum_{d=3,5,...}(k_{AF}^{(d)})^{0}_{\ \alpha_{1},\alpha_{2},...,\alpha_{(d-3)}}t^{\alpha_{1}}t^{\alpha_{2}}...t^{\alpha_{(d-3)}}|{\bf k}|^{d-3}\cr\cr
(\hat{\bf k}_{AF}^{(d)})\rightarrow({\bf k}_{AF})&=&
\sum_{d=3,5,...}({\bf k}_{AF}^{(d)})_{\alpha_{1},\alpha_{2},...,\alpha_{(d-3)}}k^{\alpha_{1}}...k^{\alpha_{(d-3)}}\cr\cr
&=&\sum_{d=3,5,...}({\bf k}_{AF}^{(d)})_{\alpha_{1},\alpha_{2},...,\alpha_{(d-3)}}t^{\alpha_{1}}t^{\alpha_{2}}...t^{\alpha_{(d-3)}}|{\bf k}|^{d-3} \,
\end{eqnarray}
where we have defined $t^{\alpha_{i}}=(1-\delta^{\ \alpha_{i}}_{1})(1-\delta^{\ \alpha_{i}}_{2})$.

For future convenience, we also define $({k}_{AF})^{\mu}=\Bigl(({k}_{AF})^{0},(\hat{\bf k}_{AF}^{(d)})\Bigl)$.

By combining (\ref{DHP}), (\ref{propsol}), (\ref{zxc1}) and (\ref{xcv2}) and using
then in eqs.(\ref{eqprop}), we are led to:
\begin{eqnarray}
\label{asd3}
\sum_{i,j=1}^{3}ik^{i}\varepsilon_{P}^{ij}e^{j}&=&({\bf k}_{AF})\cdot{\bf b}\ ,\cr\cr
i{\bf k}\cdot{\bf b}&=&0\ ,\cr\cr i{\bf k}\times{\bf e}&=&i\omega{\bf b}\ ,\cr\cr
\sum_{j,k,\ell=1}^{3}i\epsilon^{ijk}k^{j}(\mu_{P}^{-1})^{k\ell}b^{\ell}&=&
\sum_{j=1}^{3}-i\omega\varepsilon_{P}^{ij}e^{j}+(k_{AF})^{0}b^{i}
-\sum_{j,k=1}^{3}\epsilon^{ijk}(k_{AF})^{j}e^{k}\ ,\cr
&\ &\ 
\end{eqnarray}
where $\epsilon^{ijk}$ is the Levi-Civita 3-tensor with,
$\epsilon^{123}=1$.

From the first and second equations in (\ref{asd3}), we conclude
that
\begin{eqnarray}
\label{asd4} e^{3}=\frac{({\bf k}_{AF})\cdot{\bf b}}{ik}\ \ , \ \ b^{3}=0\ ,
\end{eqnarray}
where terms of order $R{\left| ({\bf k}_{AF}) \right|}$ were discarded.
These results show that the propagating magnetic field is
perpendicular to the direction of the propagating waves, but it does
not necessarily happen to be orthogonal to the propagating electric
field; the latter actually develops a component along $({\bf k}_{AF})$.
This effect is strictly due to the Lorentz- symmetry breaking term
in Lagrangian (\ref{modelo}).

It remains to be found an explicit form for the propagating magnetic
field as a function of $({k}_{AF})^{\mu}$. Using the fact that ${\bf
k}=k{\hat z}$, the third Eq. (\ref{asd3}) leads to
\begin{equation}
\label{asd5} {\bf b}=\frac{k}{\omega}{\hat z}\times{\bf e}\ .
\end{equation}
Substituting this relation into the first Eq.(\ref{asd4}), we have
\begin{equation}
\label{asd6} e^{3}=\frac{1}{i\omega}({\bf k}_{AF})\cdot({\hat z}\times{\bf
e})\ \rightarrow\ e^{3}= \frac{1}{i\omega}\Bigl(({k}_{AF})^{2}e^{1}-({k}_{AF})^{1}e^{2}\Bigr)\ .
\end{equation}
From Eqs (\ref{zxc1}), the last Eq. (\ref{asd3}), the second Eq.
(\ref{asd4}), Eqs (\ref{asd5}) and (\ref{asd6}), and by neglecting
terms of order $S({k}_{AF})^{\mu}$ and $R({k}_{AF})^{\mu}$, we get a rather simple
system,
\begin{eqnarray}
\label{zxc2} [(\omega^{2}-{\bf
k}^{2})+kSB_{0}^{2}]e^{1}-ik\Bigl(({k}_{AF})^{0}-({k}_{AF})^{3}\Bigr)e^{2}&=&0, \cr\cr
ik\Bigl(({k}_{AF})^{0}-({k}_{AF})^{3}\Bigr)e^{1}+[(\omega^{2}-{\bf k}^{2})+k(4RB_{0}^{2})]e^{2}&=&0\ .
\end{eqnarray}

Obviously, the system above has non-trivial solutions if, and only
if, the determinant of the corresponding matrix vanishes. For a
given wavelength, $\lambda$, the modulus of the wave vector,
$k=2\pi/\lambda$, is determined and the condition of vanishing
determinant for the coefficients matrix in (\ref{zxc2}), together
with relation (\ref{Parametros}), yields the frequencies below:
\begin{equation}
\label{zxc3}
\omega_{\pm}=k\Biggl[1-\frac{1}{4}\Biggl(11RB_{0}^{2}\pm\sqrt{9R^{2}B_{0}^{4}+
\frac{4\Bigl(({k}_{AF})^{0}-({k}_{AF})^{3}\Bigr)^{2}}{{\bf k}^{2}}}\Biggr)\Biggr],
\end{equation}
from which the corresponding phase and group velocities follow:
\begin{eqnarray}
\label{velocidades} V_{\pm}&=&\frac{\omega_{\pm}}{k}
=1-\frac{1}{4}\Biggl(11RB_{0}^{2}\pm\sqrt{9R^{2}B_{0}^{4}+\frac{4\Bigl(({k}_{AF})^{0}-({k}_{AF})^{3}\Bigr)^{2}}{{\bf k}^{2}}}\Biggr)\ ,
\end{eqnarray}
\begin{eqnarray}
\label{freq2b}
V_{g\pm}=\frac{d\omega_{\pm}}{dk}&=&V_{\pm}+
\frac{\Bigl(({k}_{AF})^{0}-({k}_{AF})^{3}\Bigr)^2}{|{\bf k}|^{3}}\frac{1}{\sqrt{9R^{2}B_{0}^{4}+
\frac{4[({k}_{AF})^{0}-({k}_{AF})^{3}]^{2}}{{\bf k}^{2}}}}\Biggl[\pm1\cr\cr
&\mp&\frac{|{\bf k}|}{[({k}_{AF})^{0}-({k}_{AF})^{3}]}\sum_{d=5,7,...}t^{\alpha_{1}}t^{\alpha_{2}}...t^{\alpha_{(d-3)}}(d-3)|{\bf k}|^{d-4}\cr\cr
&\ &\qquad \times\Bigl[(k_{AF}^{(d)})^{0}_{\ \alpha_{1},\alpha_{2},...,\alpha_{(d-3)}}-(k_{AF}^{(d)})^{3}_{\ \alpha_{1},\alpha_{2},...,\alpha_{(d-3)}}\Bigr]\Biggr]
\ . 
\end{eqnarray}

With the results (\ref{zxc3}), (\ref{velocidades}) and
(\ref{freq2b}), we are ready to carry out numerical estimates for
the corrections induced by the non-linearity and the
Lorentz-symmetry breaking parameters; the latter here show up in the
combination $[({k}_{AF})^{0}-({k}_{AF})^{3}]$.

Let us start by taking $(k_{AF}^{(d)})_{\mu}=0$ for $d>3$ in (\ref{defkaf}) and make some estimates and draw some conclusions in this case. As stated in reference \cite{AM10}, in this situation the model (\ref{defkaf}) reduces to the well known Field-Jackiw model. Also, the second and third lines of (\ref{freq2b}) disappears. With the present typical values for
the $({{k}_{AF}^{(3)})^{0}}^{\mu}$-components ($\lesssim 10^{-43} GeV$) \cite{Astro}
and for intergalactic magnetic fields ( $\sim 10^{-9}T \sim 10^{-18}
MeV^{2}$), non-linear and $({k}_{AF}^{(3)})$-Lorentz-symmetry breaking corrections are
of a comparable order of magnitude for wavelengths in the $\gamma$ -
ray region of the spectrum ($\lambda\sim 10^{-13}$ m). However, in
this situation, both effects are not of a measurable size, since
$RB_{0}^{2}\sim 10^{-41}$ ($RB_{0}^{2}$ is dimensionless). But, if we consider $({k}_{AF}^{(5)})$-effects, and also consider the bound $|({k}_{AF}^{(5)})|\leq10^{-32} Gev^{-1}$ \cite{AM10}, Lorentz-symmetry breaking effects are still very tiny, but they dominate. Actually, based on Eqs. (\ref{zxc3})-(\ref{freq2b}), we estimate the $({k}_{AF}^{(5)})$-effects to be of order of $10^{-16}$. So, in this situation, we may have Lorentz breaking effects stronger than non-linearity corrections to the group velocities.

By considering another range of magnetic fields
($\sim10^{7}T\sim10^{-2}$ $Me{V^2}$), non-linearity corrections
($RB_{0}^{2}$) are of the order of $10^{-9}$. In this case, the
Lorentz-symmetry breaking effects are, however, much smaller
($\sim10^{-40}$ in the $\gamma$-ray region, and $10^{-27}$ in the
microwave region) and then, practically undetectable. From our
study, we understand that in the range of physically realizable
magnetic fields (the ones produced in quark-gluon plasmas, by
neutron stars or by magnetars), effects of nonlinearity, whenever
detectable, are strongly dominating over the $({k}_{AF}^{(3)})$-Lorentz-symmetry
breaking corrections all over the electromagnetic spectrum. On the other hand, we have also identified a situation where $({k}_{AF}^{(5)})$-effects may dominate over non-linearity as mentioned above.

From the results (\ref{zxc3}), (\ref{velocidades}) and
(\ref{freq2b}), we also understand that detectable corrections
induced by the breaking of Lorentz covariance would be present at
very high energy scales, whenever string effects trigger the
violation of relativistic invariance and the violating parameters
may acquire values of the order of $10^{18}GeV$. However, at this
regime, a new physics is at work and the Lorentz-symmetry breaking
may induce other corrections to Maxwell equations which we are not
contemplated here.

It can also be seen from the system (\ref{zxc2}) that each mode has only
a single degree of freedom. By choosing $e_{\pm}^{1}$ as the free
variable, system (\ref{zxc2}) can be solved with (\ref{zxc3}) and
using (\ref{asd6}). Defining $\Delta_{\pm}=\omega_{\pm}^{2}-k^{2}$,
the modes ${\bf e}_{\pm}=(e_{\pm}^{1},e_{\pm}^{2},e_{\pm}^{3})$ for
the electric field turn out to be
\begin{eqnarray}
\label{eletrico} e_{\pm}^{2}&=&e_{\pm}^{1}\ \frac{\Delta_{\pm}+{\bf
k}^{2}SB_{0}^{2}}{ik[({k}_{AF})^{0}-({k}_{AF})^{3}]}\ ,\cr\cr
e_{\pm}^{3}&=&e_{\pm}^{1}\
\frac{1}{ik}\Biggl(({k}_{AF})^{2}-({k}_{AF})^{2}\frac{\Delta_{\pm}+ {\bf
k}^{2}SB_{0}^{2}}{ik[({k}_{AF})^{2}-({k}_{AF})^{1}]}\Biggr)\ .
\end{eqnarray}
The $ e_{\pm}^{3}$-components are exclusively due to the breaking of
Lorentz symmetry, but they are not independent, for they are
proportional to the $ e_{\pm}^{1}$-components. The propagating
magnetic fields are given by (\ref{asd5}), (\ref{zxc3}) and
(\ref{eletrico}) as,
\begin{eqnarray}
\label{magnetico} {\bf
b}_{\pm}=\Biggl[1+\frac{1}{4}\Biggr(11RB_{0}^{2}\pm\sqrt{9R^{2}B_{0}^{4}+
\frac{4\Bigl(({k}_{AF})^{0}-({k}_{AF})^{3}\Bigr)^2}{{\bf k}^{2}}}\Bigg)\Biggr]
(e_{\pm}^{1}{\hat y}-e_{\pm}^{2}{\hat x})\ .
\end{eqnarray}

\section{Final Remarks}

In conclusion, we have worked out the simplest solutions
for the set of dynamical equations describing electromagnetic
process in a Lorentz-symmetry breaking scenario taking into account
QED non-linear induced effects, up to the first order in two
parameters, in presence of an external constant magnetic
field.

This has been used to reach a possibly useful expression to analyze
the dispersion phenomena from astrophysical data coming from
electromagnetic waves that have passed through a region where strong
magnetic fields can be found, as in the surroundings of neutron
stars or magnetars, since extreme precision can be reach using
astrophysical sources \cite{Jackiw,dados,Astro}. In these regions
even the bending of light by magnetic fields dominates the
gravitational bending \cite{LK}.

Peculiar polarization states for the photon field have also been
found. It turns out  that the photon acquires two distinct modes,
each one with a single degree of freedom, when passing through a
region of constant strong magnetic field in a Lorentz-violating
background according to (\ref{eletrico}). By virtue of the breaking
of Lorentz symmetry, the polarization modes indicate that the photon
behaves as if it splits into independent scalars. Longitudinal
polarization, like the characteristic  $e_{\pm}^{3}$ in
(\ref{eletrico}), has been object of interest in optics for a long
time \cite{CH}. Because longitudinal components occur at the focal
region of tightly focused laser beams a lot of optical techniques
has been developed to study them, (see for instance \cite{Lee,Wang}
and references cited therein). In turn this can be useful to set
upper bounds at laboratory scale of possible Lorentz-symmetry
breaking models, in addition to the analysis of camouflage
coefficients \cite{PM} that control this kind of violation in the
SME.

\section{Acknowledgments}

The authors would like to thank CAPES, CNPq and FAPEMIG (Brazilian
agencies) for invaluable financial support. P. G. was partially
supported by Fondecyt (Chile) grant 1080260. They also express their gratitude to the Referee of the paper for the very pertinent comments and suggestions.

\end{document}